\documentclass[12pt]{article}

\usepackage{graphicx}
\usepackage{amsmath}
\usepackage{amssymb}
\usepackage{dsfont}
\usepackage{hyperref}

\usepackage{caption}
\usepackage{titlesec}
\titleformat{\section}
  {\normalfont\fontsize{15}{18}\bfseries}{\thesection}{1em}{}
\newcommand{\poiss}[1]{\left\{#1\right\}}

\newcommand{\hil}[0]{\mathcal{H}}

\begin{document}

${}$\\
\begin{center}
\vspace{36pt}
{\Large \bf Spherically symmetric solutions \\ \vspace{12pt} 
of the $\lambda$-R model}
\vspace{48pt}

{\sl R. Loll and L. Pires}\footnote{emails: r.loll@science.ru.nl, l.pires@science.ru.nl}

\vspace{24pt}

Radboud University, \\
Institute for Mathematics, Astrophysics and Particle Physics, \\
Heyendaalseweg 135, NL-6525 AJ Nijmegen, The Netherlands.\\ \vspace{2pt}

\vspace{72pt}

\end{center}

\begin{center}
{\bf Abstract}
\end{center}
We derive spherically symmetric solutions of the classical $\lambda$-R model, a minimal, anisotropic modification of
general relativity with a preferred foliation and two local degrees of freedom.
Starting from a $3+1$ decomposition of the four-metric in a general
spherically symmetric ansatz, we perform a phase space analysis of the reduced model. We show that its
constraint algebra is consistent with that of the full $\lambda$-R model, and also yields  
a constant mean curvature or maximal slicing condition as a tertiary constraint. 
Although the solutions contain the standard Schwarzschild geometry for the general relativistic value $\lambda\! =\! 1$
or for vanishing mean extrinsic curvature $K$, they are in general non-static, incompatible with asymptotic
flatness and parametrized not only by a conserved mass. We show by explicit computation that 
the four-dimensional Ricci scalar of the solutions is in general time-dependent and nonvanishing.

\newpage

\section{Introduction}
When trying to describe physics in a way that encompasses both general relativity and quantum theory, one is confronted
with the fact that in each of them ``time" seems to play a very different role. While standard quantum (field) theory on flat space
has, up to Poincar\'e transformations, a distinguished notion of time that forms part of its fixed background structure, there is no a priori 
distinguished time in general relativity, and any physical notion of time is subject to the dynamics of gravity. 

While much has been said about the role time may play in a quantum theory of gravity and on how to implement particular proposals 
in technical terms \cite{CIsh}, the primary focus of the present piece of work is classical, although its motivation stems in part from
quantum considerations. More specifically, we will investigate properties of a one-parameter family of theories of ``neighbours" of general
relativity, where a preferred foliation of spacetime by spatial hypersurfaces -- and therefore a preferred class of times -- forms part of the 
theory's background structure. 

In ordinary gravity, the fact that a Lorentzian spacetime $\cal M$ is diffeomorphic to a product 
\begin{equation}\label{fol}
{\cal M}=\mathds{R}\times \Sigma
\end{equation}	 
of a smooth spacelike hypersurface $\Sigma$ and a time direction $\mathds{R}$ follows from the usual requirement of global hyperbolicity,
which ensures the causal structure of spacetime is sufficiently well behaved. There are of course infinitely many ways of foliating
any particular spacetime geometry. If the spacetime is given by a Lorentzian metric $g_{\mu\nu}(x)$ solving the Einstein equations, different 
foliations will correspond to different choices of time coordinate. However, arbitrary spacetime-dependent 
reparametrizations of time are examples of spacetime 
diffeomorphisms, which leave the theory invariant and lead to physically equivalent metrics. In this sense, different choices of foliation can be 
considered part of the gauge freedom of general relativity.

The situation is different in the so-called $\lambda$-R model\footnote{The $\lambda$ in the ``$\lambda$-R model"
refers to a dimensionless coupling
constant in the kinetic part of the Lagrangian, while the potential part consists merely of the Ricci scalar R of the three-dimensional
slices \cite{BR1}, see below for further details.} we will study here. This theory generalizes the metric formulation of pure gravity
and is also formulated in terms of metric fields. 
It is based on a minimal, one-parameter modification of the classical Einstein-Hilbert action in a 3+1 decomposition and was first
investigated in a purely classical context in the 1990s \cite{DoCl}.  
More recently it has appeared as low-energy limit of a class of so-called non-projectable Ho\v rava-Lifshitz gravities \cite{hor,MH,TD,BR1,LP}.
These generalized gravity theories are associated with a {\it preferred} foliation of spacetime.  
As a consequence, they are only invariant under a subset of spacetime diffeomorphisms, namely those that 
do not change the preferred foliation. The remaining invariance group consists of three-dimensional diffeomorphisms acting independently on each leaf
$\Sigma_t$ (labelled by time $t$) and {\it space-independent} time reparametrizations. 
The most general local action of the metric fields which is at most quadratic in derivatives and invariant under this reduced symmetry group is not 
the Einstein action, but a two-parametric generalization thereof. The $\lambda$-R model can be viewed as the simplest such theory, obtained 
by setting one of the two new coupling constants to zero\footnote{This is the parameter associated with a term containing spatial 
derivatives of the lapse function.}. The remaining real parameter $\lambda$
appears inside the kinetic term of the Einstein-Hilbert action, which reduces to its standard, general relativistic form for $\lambda =1$.

We will not concern ourselves with the viability of Ho\v rava-Lifshitz theories as possible candidate theories of quantum gravity, but will
investigate the classical $\lambda$-R theory on its own merits, and explore some of the consequences of the altered status of time and of the 
symmetry structure, compared with usual gravity. What makes the $\lambda$-R model particularly attractive is the fact that 
it has two local physical degrees of freedom, matching those of general relativity, albeit with a second-class instead of a first-class 
constraint algebra \cite{TD,BR1,LP}. 

Building on our earlier work \cite{LP}, we will in this paper treat the case of spherical symmetry, and will look for 
a physical signature that distinguishes the $\lambda$-R gravity theory from general relativity proper. 
Within the larger context of Ho\v rava-Lifshitz gravities, results on spherically symmetric solutions and black holes have been obtained
\cite{BH1,BH2,BH3,BH4}, but to our knowledge have no overlap with our current investigation. Their new features arise from
higher-derivative operators or terms in the action depending on the vector $\nabla_i\log N$, neither of which we consider. In addition,
they assume either staticity, vanishing shift or asymptotic flatness, in contrast with crucial features exhibited by our solutions,
as we will demonstrate.

This article is organized as follows. In the remainder of this section we recall essential features of the 
classical Hamiltonian formulation of the $\lambda$-R model and some earlier results on its constraint structure, and
explain how we implement spherical symmetry. In Sec.\ \ref{red mod} we follow the metric ansatz of
\cite{KK}, including a nonvanishing radial shift, to set up a Hamiltonian formulation of the spherically symmetric sector of
the $\lambda$-R model and derive its total Hamiltonian. We determine the 
constraint algebra systematically \`a la Dirac \cite{Dir,Sund,HT}, 
solve the constraints explicitly, and compute the time evolution of the canonical variables. In Sec.\ \ref{prop} we investigate
the properties of the solutions. This also involves a careful discussion of the fall-off conditions of the dynamical variables.
We derive explicit expressions for the extrinsic curvature, the spacetime metric and the four-dimensional scalar curvature of
the solutions, and write them in a form that can be compared directly with general relativity in a CMC formulation. 
For $\lambda\!\not=\! 1$ and foliations with nonvanishing mean curvature, the solutions cannot be interpreted as solutions
to the vacuum Einstein equations. Finally, in Sec.\ \ref{conclusion} we present a summary and conclusion of our results.

\subsection{Hamiltonian set-up}
\label{hamil}

Before getting to the specifics of how we implement spherical symmetry, let us recall some important ingredients of the general
set-up. Because of the presence of the preferred spatial foliation\footnote{We will consider a mild generalization in the main part of the paper, 
where we allow the space-like leaves of the foliation to become null asymptotically; we will continue to refer to such leaves as ``spatial
hypersurfaces".} and since we will work within the Hamiltonian formalism throughout, we use the $3\! +\! 1$ ADM decomposition of the metric \cite{adm}
with line element
\begin{equation}
ds^2=-N^2 dt^2+g_{ij}(dx^i+N^i dt)(dx^j+N^jdt),
\label{lineelement}
\end{equation}
where $N(x)$ is the lapse function and $N^i(x)=g^{ij}N_j(x)$ the shift three-vector. The inverse of the spatial metric $g_{ij}$ is denoted
by $g^{ij}$, not to be confused with the spatial components of the inverse four-metric $g^{\mu\nu}$.
We will reconstruct the full four-dimensional metric $g_{\mu\nu}$ only when we have obtained a class of spherically symmetric solutions of the theory. 
In the ADM-setting, the usual Einstein-Hilbert action on a differential manifold $\cal M$ of the form (\ref{fol}) is given by
	\begin{subequations}\label{EH}\begin{align}
	S=&\,\kappa\int dt\int d^3x\sqrt{g}\, N \left(G^{ijkl}K_{ij}K_{kl}+{\cal R}\right)\label{Gmetric}\\
	=&\,\kappa\int dt \int d^3x \sqrt{g}\, N\left(K^{ij}K_{ij}-K^2+{\cal R}\right),
	\end{align}\end{subequations}
where $\kappa$ depends on Newton's constant $G$ through $\kappa\! =\!\frac{1}{16\pi G}$, 
$g$ denotes the determinant of the three-metric $g_{ij}$, ${\cal R}$ is the three-dimensional Ricci scalar on $\Sigma_t$, 
and $K_{ij}$ the extrinsic curvature
	\begin{equation}
	K_{ij}=\frac{1}{2N}\left(\dot{g}_{ij}-\nabla_iN_j-\nabla_jN_i\right).
	\label{extr}
	\end{equation}
The overdot in definition (\ref{extr}) denotes a time derivative, and $\nabla_i$ is the covariant derivative with respect to $g_{ij}$. 
Lastly, in writing the action \eqref{Gmetric} we employed the Wheeler-DeWitt metric $G^{ijkl}$ \cite{BdW,DG},
	\begin{equation}
	G^{ijkl}:=\frac{1}{2}\left(g^{ik}g^{jl}+g^{il}g^{jk}\right)-g^{ij}g^{kl},
	\end{equation}
which defines a metric on the infinite-dimensional space of three-metrics on $\Sigma$.

The expression $G^{ijkl}K_{ij}K_{kl}$ in \eqref{Gmetric} is the most general local and spatially covariant term which is of second order 
in time derivatives. By contrast, we want to restrict ourselves to foliation-preserving diffeomorphisms, whose infinitesimal generators are
	\begin{equation}
	\label{repar}
	\delta t=f(t),\qquad \delta x^i=\zeta^i(x,t),
	\end{equation}
and act on the ADM fields according to
	\begin{subequations}\label{fp diff}\begin{align}
	&\delta g_{ij}=\zeta^k\partial_kg_{ij}+f\dot{g}_{ij}+\left(\partial_i\zeta^k\right)g_{jk}
	+\left(\partial_j\zeta^k\right)g_{ik}, \\
	&\delta N_i=\left(\partial_i\zeta^j\right)N_j+\zeta^j\partial_jN_i
	+\dot{\zeta}^jg_{ij}+\dot{f}N_i+f\dot{N}_i, \\
	&\delta N=\zeta^j\partial_jN+\dot{f}N+f\dot{N}.
	\end{align}\end{subequations}
Under the transformations \eqref{fp diff}, the two scalars $K^{ij}K_{ij}$ and $K^2$ are independently invariant and we are therefore
free to introduce a new dimensionless relative coupling constant $\lambda$ between them. The resulting action reads
	\begin{subequations}\label{lamR}\begin{align}
	S=&\,\kappa \int dt\int d^3x \sqrt{g}\, N\left(K_{ij}K^{ij}-\lambda K^2+{\cal R}\right)\\
	=&\,\kappa\int dt\int d^3x\sqrt{g}\, N\left(G^{ijkl}_{\lambda} K_{ij}K_{kl}+{\cal R}\right),
	\end{align}\end{subequations}
where $G^{ijkl}_{\lambda}$ now denotes a {\it generalized} Wheeler-DeWitt metric of the form
	\begin{equation}\label{genWdw}
	G^{ijkl}_{\lambda}=\frac{1}{2}\left(g^{ik}g^{jl}+g^{il}g^{jk}\right)-\lambda g^{ij}g^{kl}.
	\end{equation}
The fact that (\ref{genWdw}) constitutes the most general ultralocal metric (a metric not involving derivatives of $g_{ij}$) on 
the space of metrics motivated the earlier-mentioned study \cite{DoCl} of the generalized gravity theory given by (\ref{lamR}). 
Since the generalized Wheeler-DeWitt
metric does not have an inverse for $\lambda\! =\! 1/3$, we will not consider this special value. 
In addition, as will become apparent in Sec.\ \ref{solve}, we must exclude values $\lambda <1/3$ to obtain physically
sensible solutions. In the remainder of this paper we will therefore restrict ourselves to the parameter range $\lambda >1/3$.

With the action (\ref{lamR}) we have finally arrived at the ``$\lambda$-R model'', a term coined originally 
in \cite{BR1}, whose authors were the first to study the Hamiltonian formulation of the model for asymptotically flat spatial hypersurfaces. 
The generalization of this analysis to closed and compact 
hypersurfaces was performed in \cite{LP}, with results that turn out to be applicable also to the case of asymptotically null hypersurfaces 
considered below. 

In this work, we will determine the solutions of the $\lambda$-R model without a cosmological constant, as defined by the action \eqref{lamR}, 
for the case that the leaves of the foliation possess an additional spherical symmetry. 
For $\lambda\neq 1$ the constraint algebra becomes second class, due to the appearance of the second-class constraint 
	\begin{equation}\label{CMC}
	\nabla_i\pi=0,
	\end{equation}
where $\pi\! =\! g_{ij}\pi^{ij}$ is the trace of the three-momentum $\pi^{ij}$ conjugate to $g_{ij}$.\footnote{With the sign convention \eqref{extr} 
for the extrinsic curvature, the relation between $\pi$ and the trace $K:=g^{ij}K_{ij}$ of the extrinsic curvature is given by
$\pi=-\sqrt{g}(3\lambda -1)K$.} 
Eq.\ \eqref{CMC} is a ``constant mean curvature (CMC) condition", and familiar from standard general relativity,
where it can be adopted as a possible gauge choice to gauge-fix the Hamiltonian constraint. By contrast, in the $\lambda$-R model
it appears as a second-class constraint during the Dirac constraint analysis, where together with the Hamiltonian constraint
${\cal H}_\lambda\approx 0$ it forms a pair of second-class constraints. 
Requiring (\ref{CMC}) to be conserved in time in general relativity in CMC gauge \cite{JY1,JY2,JY3,YM1,YM2}
leads to a consistency condition, which has a direct ($\lambda$-dependent) analogue in the $\lambda$-R model, obtained by demanding
closure of its constraint algebra \cite{LP}.
 
In the context of general relativity, condition \eqref{CMC} has been studied and shown to have solutions for open hypersurfaces
that are asymptotically flat or asymptotically null, and for closed and compact hypersurfaces. 
With asymptotic flatness, the only consistent solution is $\pi=0$ (known in general relativity as ``maximal slicing condition"), 
the asymptotically null case requires 
$\pi=\sqrt{g} A(t)$, with a function $A(t)$ that is nowhere vanishing\footnote{If there is a time $t$ for which $A$ is zero,
the corresponding hypersurface will no longer be asymptotically null.}, and the compact case allows for any $A(t)$.  

While in the case of gravity, these are mere gauge choices that do not affect the physical content of solutions, the
situation in the $\lambda$-R model is different. An analysis of the initial value problem of the model \cite{LN} 
shows that different choices of the function $A(t)$ are in general associated with physically
inequivalent solutions. Moreover, for $\pi\not= 0$ they can no longer be thought of as (gauge-fixed) solutions to general
relativity and therefore genuinely transcend that theory. One finds that the $\lambda$-dependence in these cases
cannot be absorbed simultaneously in the initial data and the evolution equations by suitable redefinitions of constants or
dynamical variables \cite{LN}.

The general phenomenon just described will be illustrated by our discussion of the spherically symmetric case. 
In our analysis below we also find a $\lambda$-dependence 
when solving the equations of motion. As we will show explicitly, this implies a nontrivial generalization of the CMC slicings of
the Schwarzschild solution in ordinary gravity \cite{Ise,CMCSch1,CMCSch2}. The general form of a CMC foliation of the Schwarzschild
spacetime was obtained relatively recently \cite{CMCSch}, while similar constructions for the Kruskal extension 
and more general black holes are the subject of ongoing research \cite{rnCMC,recSch,recSch2}.

\subsection{Spherical symmetry}
\label{spher cons}

A spacetime $({\cal M},g_{\mu\nu})$ is spherically symmetric if the group $SO(3)$ acts on it as a group of isometries and if its orbits
are spacelike two-spheres. Physically this is relevant in the presence of an isolated spherically symmetric source.
To determine the gravitational field outside it,
one solves the vacuum Einstein equations in a spherically symmetric ansatz. In general relativity, one can always
choose coordinates such that all spherical orbits lie in hypersurfaces of constant time. The Killing vectors generating the
$SO(3)$-isometries are then everywhere tangent to the constant-time slices. 

By contrast, the $\lambda$-R model comes with a {\it preferred} foliation into slices $\Sigma_t$ of constant time. The slices can be
relabelled, but by assumption the foliation itself cannot be changed by any of the allowed diffeomorphisms. As a consequence, one can distinguish
between two realizations of spherical symmetry: either all $SO(3)$-orbits are contained in leaves of the preferred foliation, or they are not.
Our analysis below will deal with the former case, where the orbits are ``aligned" with the preferred foliation, and which is
technically simpler. There is no obvious reason for not also considering the more general, non-aligned case, but we will not do so here.

Another way of phrasing the issue is to assume the presence of an isolated point-like source somewhere in a given,
foliated spacetime described by $\lambda$-R gravity. Because of the absence of space-dependent time reparametrizations,
there will in general be an obstruction to performing a coordinate transformation that eliminates cross terms proportional
to $dt\!\; d\phi$ and $dt\;\! d\theta$ in the line element to arrive at a block-diagonal metric of the form 
\begin{equation}
ds^2=d\tau^2(t,r)+R^2(t,r)d\Omega^2(\theta,\phi),
\label{standard}
\end{equation} 
where $d\tau^2$ describes an indefinite two-surface and $\theta$ and $\phi$
are standard angular coordinates on the unit two-sphere with line element $d\Omega^2=d\theta^2+\sin^2\theta \; d\phi^2$. 

Even when the $SO(3)$-orbits are contained in the leaves of the preferred foliation and the metric is of
the form (\ref{standard}) -- as we will be assuming --
it will in general not be possible to eliminate the cross term proportional to $dt\;\! dr$ in $d\tau^2$ and
thus rewrite the line element as
\begin{equation}
ds^2=-a^2(t,r)dt^2+b^2(t,r)dr^2+R^2(t,r)d\Omega^2(\theta,\phi).
\label{cross}
\end{equation}
However, note that eliminating the radial shift 
forms part of a standard derivation of Birkhoff's theorem in general relativity (see, for example,
\cite{YCB}), according to which the (static and asymptotically flat) Schwarzschild metric is the unique solution of the vacuum
Einstein equations outside a nonrotating, spherically symmetric gravitating body. This raises the question whether there
is an analogue of Birkhoff's theorem for the $\lambda$-R model. 
One reason to expect that the Schwarzschild solution may have to be generalized is the fact that 
transforming the Schwarzschild metric from standard Schwarzschild coordinates to a CMC slicing with $\pi\neq 0$ in
general relativity requires a space-dependent time reparametrization \cite{CMCSch1,CMCSch2,CMCSch}. 
In standard gravity this is a particular four-dimensional diffeomorphism, which implies that the metrics before and
after the transformation are physically equivalent. This can no longer be the case in the $\lambda$-R
model, where this type of diffeomorphism is not allowed because it does not preserve the foliation. The consequences
of this observation will be analyzed in the following sections.

\section{Reduced model and phase space analysis}
\label{red mod}
From now on, we will assume that the geometry on each leaf $\Sigma_t$ of the foliation is spherically symmetric in the sense that $SO(3)$-orbits 
through points in $\Sigma_t$ never leave $\Sigma_t$, and are therefore aligned with the foliation, as described in the previous section. 
Since the $\lambda$-R model is invariant under spatial diffeomorphisms, we can without loss of generality write the spatial line element of 
$\Sigma_t$ as 
	\begin{equation}\label{sLin}
	dS^2=\mu^2(t,r)dr^2+R^2(t,r)d\Omega^2,
	\end{equation}
for functions $\mu(t,r)$ and $R(t,r)$ that are everywhere positive.
Taking \eqref{sLin} as a starting point, we follow the treatment of \cite{KK} in writing the line element of the four-dimensional spacetime as
	\begin{equation}\label{5}
	ds^2=-\left(N^2-\mu^2\xi^2\right)dt^2+2\mu^2\xi drdt+\mu^2dr^2+R^2d\Omega^2
	\end{equation}
for real functions $\xi(t,r)$ and $N(t,r)>0$. Dotted and primed quantities will denote partial derivatives with respect to $t$ and $r$ 
respectively. Under transformations of $r$, $R$ behaves like a scalar and $\mu$ like a scalar density of weight 1. 
This implies that $\xi$ is a scalar density of weight -1 and $R'(t,r)$ a density of weight 1. 
While $N$ and $\xi$ are absent from \eqref{sLin}, we will not treat them as Lagrange multipliers but as fields.

Having set up the metric variables, we can compute both the intrinsic scalar curvature $\cal R$ of $\Sigma_t$,
	\begin{equation}
	{\cal R}=\frac{2}{R^2}\left(1-\frac{\left(R'\right)^2}{\mu^2}-2\frac{R}{\mu}\left(\frac{R'}{\mu}\right)'\right),
	\label{intrR}
	\end{equation}
and the extrinsic curvatures
	\begin{subequations}\begin{align}
	&K_{rr}=\frac{1}{N}\left(\mu\dot{\mu}-\mu^2\xi'-\mu\mu'\xi\right) ,\label{Krr}\\
	&K_{\theta\theta}=\frac{1}{N}\left(R\dot{R}-RR'\xi\right) =\frac{K_{\phi\phi}}{\sin^2\theta}.\label{Kpp}
	\end{align}\end{subequations}
Noting that the determinant satisfies $\sqrt{g}=\mu R^2\sin\theta$, we can now integrate out the angular dependence of the action,
	\begin{subequations}\label{8}\begin{align}
	S=&\;\kappa\int dt\int_{-\infty}^{+\infty} dr\int_0^{\pi}d\theta\int_0^{2\pi}d\phi\,\sqrt{g}N\left(K_{ij}K^{ij}-\lambda K^2 
	+{\cal R}\right)\\
	=&\; 4\pi\kappa \int dt\int_{-\infty}^{+\infty} dr\, \mu R^2 N\left(K_{ij}K^{ij}-\lambda K^2+{\cal R}
	\right).\label{8second}
	\end{align}\end{subequations}
We have chosen the range $r\!\in\!\left(-\infty,+\infty\right)$ for the radial coordinate, which implies that the
spatial hypersurfaces run from the left to the right wedge of the Kruskal diagram.
This also matches the CMC treatment of the Schwarzschild solution \cite{CMCSch} we will be comparing with later on. 
It will turn out that our result for the special choice of a vanishing trace of the extrinsic curvature, $K\! =\! 0$, coincides 
with the Schwarzschild metric of general relativity, written in a constant mean curvature slicing. 
Potentially new spherically symmetric solutions will arise in the $\lambda$-R setting for $K\!\neq\! 0$, again in a CMC slicing.

Setting $4\pi\kappa\! =\! 1$ in what follows (this is equivalent to setting $G=\frac{1}{4}$ in units where $c=1$), we define conjugate momentum variables from the Hamiltonian form \eqref{8second} of
the action, yielding
	\begin{subequations}\begin{align}
	&\phi_N:=\frac{\delta S}{\delta \dot{N}}=0,\qquad \phi_{\xi}:=\frac{\delta S}{\delta \dot{\xi}}=0,\label{prim}\\
	&\pi_{\mu}=\frac{2R}{N}\left[\left(1-\lambda\right)\frac{R}{\mu} \label{9b}
	\left(\dot{\mu}-\mu\xi'-\mu'\xi\right)-2\lambda\left(\dot{R}-R'\xi\right)\right],\\
	&\pi_R=\frac{4\mu}{N}\left[\left(1-2\lambda\right)\left(\dot{R}-R'\xi\right)-\lambda\frac{R}{\mu}
	\left(\dot{\mu}-\mu\xi'-\xi\mu'\right)\right]\label{9c}.
	\end{align}\end{subequations}
The momenta $\phi_N$ and $\phi_\xi$ conjugate to the fields $N$ and $\xi$ define
the primary constraints of the theory. 
Inverting \eqref{9b} and \eqref{9c}, we can write the Hamiltonian, first without any primary constraints, as
	\begin{equation}\label{10}
	H=\int dr\left(\xi\hil_r+N\hil_{\lambda}\right)+H_{\partial\Sigma},
	\end{equation}
where $\hil_r$ and $\hil_{\lambda}$ stand for the phase space functions
	\begin{subequations}\begin{align}
	\label{11a}\hil_r=\; &\pi_RR'-\mu\pi_{\mu}',\\ \hil_{\lambda}=\; &
	\frac{2\lambda-1}{4\left(3\lambda-1\right)}\frac{\mu\pi_{\mu}^2}{R^2}
	+\frac{\lambda-1}{8\left(3\lambda-1\right)}\frac{\pi^2_R}{\mu}
	-\frac{\lambda}{2\left(3\lambda-1\right)}\frac{\pi_{\mu}\pi_R}{R}\\ \nonumber
	&-2\left(\mu-\frac{\left(R'\right)^2}{\mu}-2R\left(\frac{R'}{\mu}\right)'\right).
	\end{align}\end{subequations}
We have added a boundary term $H_{\partial\Sigma}$ to the Hamiltonian, which must be chosen such that the action is sufficiently
``differentiable" in the sense of Regge and Teitelboim \cite{RgTt}. Its precise form will become important once we discuss boundary and
fall-off conditions for the fields, an issue we will return to in Sec.\ \ref{fall} below.
Finally, after adding the primary constraints, the total Hamiltonian $H_{tot}$ reads
	\begin{equation}
	H_{tot}=\int dr\left\{N\hil_{\lambda}+\xi\hil_r+\alpha\phi_N+\beta\phi_{\xi}\right\}+H_{\partial\Sigma},
	\end{equation}
where $\alpha$ and $\beta$ are Lagrange multipliers.

\subsection{Constraint algebra}
\label{con alg}
Since the total Hamiltonian $H_{tot}$ of the $\lambda$-R model is linear in both the radial shift $\xi$ and the lapse $N$, 
Poisson-commuting it with the primary constraints yields the radial momentum and Hamiltonian constraints,
	\begin{subequations}\begin{align}
	&\dot{\phi}_{\xi}=\poiss{\phi_{\xi},H_{tot}}=-\hil_r\approx 0,\label{13a}\\
	&\dot{\phi}_N=\poiss{\phi_N,H_{tot}}=-\hil_{\lambda}\approx 0,
	\end{align}\end{subequations}
which must themselves be preserved in time. 
Since we still have invariance under spatial diffeomorphisms, 
requiring that the momentum constraint $\hil_r$ hold for all times yields the same expression as in general relativity, 
	\begin{equation}\label{diffeomorphism}
	\dot{\hil}_r=\poiss{\hil_r,H_{tot}}=2\hil_r\xi'+\xi\hil_r'+\hil_{\lambda}N',
	\end{equation}
which vanishes on the constraint surface, without yielding any further constraints.
Computing the time derivative of ${\hil}_{\lambda}$ results in
	\begin{equation}
	\dot{\hil}_{\lambda}=\left(\xi\hil_\lambda\right)'+\frac{2N'+N\partial_r}{3\lambda-1}\left[
	2\lambda\, \frac{\hil_r}{\mu^2}+\left(\lambda-1\right)\left(-2\frac{\pi_{\mu}}{\mu R}R'
	+\frac{R}{\mu}\Big(\frac{\pi_R}{\mu}\Big)' \right)\right],\label{15}
	\end{equation}
which vanishes in a straightforward manner only for the general relativistic value $\lambda\! =\! 1$.	
Demanding that the right-hand side of (\ref{15}) vanish on the constraint surface for arbitrary values of $\lambda$
yields a tertiary constraint, which after some algebraic manipulations takes the form
	\begin{equation}\label{16}
	\frac{R^2}{\mu}\bigg(N^2 \Big(\frac{\pi_{\mu}}{R^2}+\frac{\pi_R}{R\mu}\Big)' \bigg)' \approx 0.
	\end{equation}
It is solved by setting
	\begin{equation}\label{17}
	\omega:= \mu \pi_{\mu}+R\pi_R-A(t)\mu R^2=0,
	\end{equation}
where $A(t)$ is a function of time which we will later show to be proportional to the trace of the extrinsic curvature. 
Recall from the general analysis in \cite{LP} that the CMC condition is given by $\nabla_i\pi=0$ and solved by $\pi-A(t)\sqrt{g}=0$. 
This means that equation \eqref{17} can be viewed as the implementation of the CMC condition on the reduced phase space. 
In geometric terms, it implies the extrinsic curvature has a trace that is spatially constant.
Next, we must demand that the time derivative of (\ref{17}) vanishes on the constraint surface, $\dot{\omega}\approx 0$.
This yields a lapse-fixing equation as a quaternary constraint, namely,
	\begin{align}
	\nonumber \dot{\omega}=&\;\frac{\partial}{\partial t}\omega+\poiss{\omega,H_{tot}}=-\dot{A}\mu R^2
	+\poiss{\omega,H_{tot}}\\
	\approx &\; 4\mu R^2\left\{\left({\cal R}+\frac{A^2}{8\left(3\lambda-1\right)}-\frac{1}{\mu R^2}\partial_r\bigg(
	\frac{R^2}{\mu}\partial_r\bigg)\right)N-\frac{\dot{A}}{4}\right\}\approx 0,\label{18}
	\end{align}
where the spatial Ricci scalar $\cal R$ was given in (\ref{intrR}) above. The Dirac algorithm ends after imposing that
the lapse-fixing equation \eqref{18} should be preserved in time, which fixes the Lagrange multiplier $\alpha$ and in turn determines the 
time evolution $\dot{N}$ of the lapse function. We will do this in the same way we dealt with relation \eqref{15}, 
first solving eq.\ (\ref{18}) for $N$ and then demanding time preservation.  

\subsection{Solving the constraints}
\label{solve}
Using eq.\ \eqref{17} to eliminate $\pi_R$ from the momentum constraint $\hil_r\approx 0$ and solving the resulting differential 
equation for $\pi_\mu$ we obtain 
	\begin{equation}\label{25}
	R'\left(AR\mu -\frac{\mu\pi_{\mu}}{R}\right)-\mu\pi_{\mu}'=0\;\; \Rightarrow\;\;
	\pi_{\mu}=\frac{C}{R}+\frac{A}{3}R^2,
	\end{equation}
where $C\! =\! C(t)$ is a new constant of integration, possibly time-dependent. We will see in Sec.\ \ref{princ} that $C(t)$ is 
related to the transverse-traceless components of the extrinsic curvature tensor. 
Writing now the radial momentum $\pi_R$ as a function of $A$ and $C$,
	\begin{equation}\label{26}
	\pi_R=\mu\left(\frac{2}{3}AR-\frac{C}{R^2}\right),
	\end{equation}
we have succeeded in solving both momenta in terms of metric variables and two integration constants. 
After substituting these solutions into the Hamiltonian constraint $\hil_{\lambda}\!\approx\! 0$
and performing some algebraic manipulations, it becomes a total derivative which we can immediately solve,
	\begin{subequations}\label{21}\begin{align}
	&\left[\left(R\left(\frac{R'}{\mu}\right)^2\right)-R-\frac{C^2}{16R^3}-\frac{A^2}{72\left(3\lambda-1\right)}R^3\right]'=0\\ \Rightarrow &
	\left(R\left(\frac{R'}{\mu}\right)^2\right)-R-\frac{C^2}{16R^3}-\frac{A^2}{72\left(3\lambda-1\right)}R^3=-8m.\label{21b}
	\end{align}\end{subequations}
Its solution introduces a new integration constant, denoted by $m$. As we will show in the next section, the Schwarzschild mass $M_s$ for $A\!\neq\! 0$ is a $\lambda$-dependent 
combination of $m$, $A$ and $C$. Inverting eq.\ \eqref{21b}, we can write the metric variable $\mu$ in terms of $R$, 
its derivatives and integration constants as
	\begin{equation}\label{28}
	\frac{\mu^2}{\left(R'\right)^2}=\frac{1}{B(R)},
	\end{equation}
where we have introduced the notation $B(R)$ as a shorthand for the function
	\begin{equation}
	B\left(R;m,A,C\right)=1-\frac{8m}{R}+\frac{C^2}{16R^4}+\frac{A^2R^2}{72\left(3\lambda-1\right)}.
	\label{bform}
	\end{equation}
Note that for $A\! =\! C\! =\! 0$ and $R\! =\! r$, we recover the metric component $g_{rr}\! =\!\mu^2$ of the standard Schwarzschild 
solution with mass $M_s\! =\! 16m$.
Formula (\ref{bform}) also illustrates why $\lambda < 1/3$ leads to unphysical behaviour, as
stated in Sec.\ \ref{hamil} above. For these parameter values the function $B(R)$ 
and therefore the metric component $g_{rr}$ will become negative for sufficiently large $R$, as a result of which the hypersurface
will have the wrong signature.

Next, in order to solve the lapse-fixing equation \eqref{18}, we first determine the most general solution to the associated 
homogeneous equation. It is given by 
	\begin{equation}\label{lap}
	N=\sqrt{B}\left(n_1+\int_{r_0}^rd\tilde{r} \frac{R'}{B^{3/2}}\frac{n_2}{R^2}\right),\;\;\; \mathrm{(homogeneous\; case)}
	\end{equation}
where $n_1$ and $n_2$ are (possibly time-dependent) integration constants. We will show later that $n_1$ determines 
the behaviour of the lapse at spatial infinity and that $n_2$ measures the time derivative of the transverse-traceless components of the extrinsic 
curvature. To obtain a solution of the full, inhomogeneous equation \eqref{18}, we add to (\ref{lap}) a particular solution of
\eqref{18}, resulting in 
	\begin{equation}\label{36}
	N_{sol}=\sqrt{B}\left(n_1+\int_{r_0}^r d\tilde{r} \frac{R'}{B^{3/2}}\left(\frac{n_2}{R^2}-\frac{\dot{A}R}{12}
	\right)\right).
	\end{equation}
Having solved the lapse-fixing equation $\dot{\omega}=0$, we write the quaternary constraint induced by it as
	\begin{equation}\label{cal m}
	{\cal M}:= N-N_{sol}\approx 0.
	\end{equation}
where $N_{sol}$ refers to the right-hand side of \eqref{36}. 
To make sure the constraint \eqref{cal m} is preserved in time, we take its Poisson brackets with the Hamiltonian $H_{tot}$.
We use $H_{tot}$ in its original version, that is, without replacing $\pi_{R}$, $\pi_{\mu}$ or $\mu$ by their solutions, and will
therefore also re-express the right-hand side of eq.\ \eqref{36} as far as possible in terms of these
variables. This is a well-defined operation on the constraint surface, leading to
	\begin{equation}\label{Nsol}
	N_{sol}\approx \pm \frac{R'}{\mu}\left(n_1+\int_{r_0}^r d\tilde{r}\frac{\mu^3}{\left(R'\right)^2}\, b(R)\right),
	\end{equation}
where the plus sign must be chosen for positive $R'$ and the minus-sign for negative $R'$, to make sure that $\sqrt{B}$, defined
by taking the square-root of eq.\ (\ref{28}), is always positive (recall that by our initial assumption $\mu>0$). We will show later
that $R=|r|$ is a consistent gauge choice, which makes $N_{sol}$ of eq.\ \eqref{Nsol} well defined for $r\not= 0$.
For ease of writing, we have in eq.\ \eqref{Nsol} introduced another abbreviation, namely,
	\begin{equation}
	b(R):=\frac{n_2}{R^2}-\frac{\dot{A}R}{12}
	\end{equation}
for a combination of terms that will appear frequently below.
When computing the time development $\dot{\cal M}$ of the constraint (\ref{cal m}), we must include
explicit time derivatives because of the dependence of $N_{sol}$ on the time-dependent 
quantities $n_1$, $n_2$ and $\dot{A}$. Taking them into account, the condition for time preservation of
this constraint reads
	\begin{equation}\label{poiss M}
	\dot{\cal M}=\frac{\partial}{\partial t} {\cal M} + \poiss{{\cal M},H_{tot}}=-\frac{\partial}{\partial t} N_{sol}+\alpha - \poiss{N_{sol},H_{tot}}
	\approx 0.
	\end{equation}
After a long but unilluminating computation, the remaining Poisson bracket in \eqref{poiss M} is found to be
	\begin{align}
	\poiss{N_{sol},H_{tot}}
	=&\;\xi N' -\frac{N}{R'}\left(\frac{AR}{6\left(3\lambda-1\right)}+\frac{C}{4R^2}\right)\left(N'+\frac{b R'}{B}\right)\nonumber
	\\& + \sqrt{B} \int_{r_0}^r d\tilde{r}\; \frac{3R'b^2}{B^{5/2}}\left(\frac{AR}{6\left(3\lambda-1\right)}+\frac{C}{4R^2}\right).\label{N sol comp}
	\end{align}
To obtain eq.\ (\ref{N sol comp}), we have discarded all boundary terms evaluated at $r_0$. This is justified because we will 
in due course set $r_0=\pm\infty$, limits for which these terms vanish with the boundary conditions we will adopt later. 
Using this result, eq.\ (\ref{poiss M}) can be written as
	\begin{align}
	\alpha=&\; \sqrt{B} \int_{r_0}^rd\tilde{r}\left(\frac{1}{B^{3/2}}\left(
	R'\left(\frac{\dot{n}_2}{R^2}-\frac{\ddot{A}R}{12}\right)\right) + \frac{3R'b^2}{B^{5/2}}\left(\frac{AR}{6\left(3\lambda-1\right)}+\frac{C}{4R^2}\right)\right) \nonumber \\
	&+\sqrt{B}\;\dot{n}_1+\xi N'-\frac{N}{R'}\left(\frac{AR}{6\left(3\lambda-1\right)}+\frac{C}{4R^2}\right)\left(N'+\frac{b R'}{B}\right) \label{33}
	\end{align}
for the Lagrange multiplier $\alpha$.	
As we will see when discussing the time evolution equations, imposing $\dot{N}=\alpha$, with $\alpha$ given by \eqref{33}, yields no 
nontrivial conditions when the equations of motion for the coordinates ${g}_{ij}$ and momenta ${\pi}^{ij}$ are satisfied. 
Before determining the time evolution of the canonical variables, we will in the next subsection comment briefly
on the first- and second-class nature of the constraints. 

\subsection{Classification of the constraints}
The $\lambda$-R model with spherical symmetry has eight phase space variables, namely,
$\left(\mu,R,N,\xi,\pi_{\mu},\pi_R,\phi_N,\phi_{\xi}\right)$, and six constraints,
	\begin{equation}\label{34}
	\phi_{\xi}=0,\quad \phi_N=0, \quad \hil_r\approx 0, \quad \hil_{\lambda}\approx 0, \quad 
	\omega\approx 0,\quad {\cal M}\approx 0.
	\end{equation}
The only obvious first-class constraint of the set \eqref{34} is $\phi_{\xi}\! =\! 0$, because none of the constraints depend
on $\xi$. From our earlier computations of $\dot{\hil}_r$, $\dot{\hil}_{\lambda}$ and $\dot{\omega}$
we deduce that the radial momentum constraint $\hil_r$ Poisson-commutes with $\hil_{\lambda}$ and $\omega$, 
as well as with $\phi_{\xi}$ and $\phi_N$.
However, the constraint ${\cal M}$ does not, since by virtue of eq.\ \eqref{N sol comp}
	\begin{equation}
	\poiss{N-N_{sol}, \hil_r}= - N',
	\end{equation}
which does not vanish on the constraint surface. The point is that in its current form the constraint $\hil_r$ 
only generates spatial diffeomorphisms of $\mu$, $R$ and their conjugate momenta. This can be remedied easily by
adding to it a term linear in the constraints\footnote{The necessity to redefine ${\cal H}_r$ already arose in the context of the full model \cite{LP}.}, 
which is always allowed. The modified momentum constraint we will be
using from now on, 
	\begin{equation}
	\tilde{\hil}_r:=\hil_r+\phi_N N'\approx 0,
	\end{equation}
generates infinitesimal diffeomorphisms of the lapse and its momentum and also Poisson-commutes with $\cal M$
on the constraint surface since 
	\begin{equation}
	\poiss{N-N_{sol},\tilde{\hil}_r}\approx N'-N'=0.
	\end{equation}
At the same time, this means that $\tilde{\hil}_r\approx 0$ is first-class. To summarize, we have
two first-class constraints, 
$\tilde{\hil}_r\approx 0$ and $\phi_{\xi}=0$, and the remaining four constraints are second-class.

\subsection{Time evolution}
\label{tev}
After having determined and solved the complete constraint algebra of the system, we will now compute the time evolution
of the canonical variables. Starting with the metric variables we find
	\begin{subequations}\begin{align}
	&\dot{R}=\poiss{R,H_{tot}}=\frac{N}{4\left(3\lambda-1\right)}\left(\left(\lambda-1\right)
	\frac{\pi_R}{\mu}-2\lambda\frac{\pi_{\mu}}{R}\right)+R'\xi.\label{37b}\\
	&\dot{\mu}=\poiss{\mu,H_{tot}}=\frac{N}{2\left(3\lambda-1\right)}\left(\left(2\lambda-1\right)
	\frac{\mu\pi_{\mu}}{R^2}-\lambda\frac{\pi_R}{R}\right)+\xi'\mu+\xi\mu', \label{37a}
	\end{align}\end{subequations}
Substituting the expressions for the canonical momenta $\pi_\mu$ and $\pi_R$ from eqs.\ (\ref{25}), (\ref{26}) into
eq.\ \eqref{37b}, we determine the radial component of the shift as 
	\begin{equation}\label{44}
	\xi=\frac{\dot{R}}{R'}+\frac{N}{R'}\left(\frac{C}{4R^2}+\frac{AR}{6\left(3\lambda-1\right)}
	\right).
	\end{equation}
Using in addition the solutions for $\mu^2$, $N$ and $\xi$, obtained in eqs.\ \eqref{28}, \eqref{36} and \eqref{44} respectively, and
substituting everything into the expression \eqref{37a} for $\dot\mu$ gives
	\begin{equation}
	\frac{R'}{BR}\left\{\left(\frac{n_2A}{3\left(3\lambda-1\right)}-\frac{C\dot{A}}{24}-8\dot{m}
	\right)+\frac{C}{2R^3}\left(\frac{\dot{C}}{4}+n_2\right)\right\}=0,
	\end{equation}
whose solution is
	\begin{equation}
	8\dot{m}=\frac{n_2A}{3\left(3\lambda-1\right)}-\frac{C\dot{A}}{24} \quad \land\quad  
	\left(\dot{C}=-4n_2 \quad \lor\quad C=0\right).
	\end{equation}
The relation $n_2\! =\! -\frac{\dot{C}}{4}$ implies that $n_2$ measures the time evolution of the transverse-traceless 
components of the extrinsic curvature captured by $C$, 
while the relation involving $\dot{m}$ will later be used to define a conserved quantity $M$, related to the mass. 
Note that $\dot{m}\! =\! 0$ holds for $A\! =\! 0$ and for the special case $\dot{C}\! =\!\dot{A}\! =\! 0$.

The equations of motion for the canonical momenta $\pi_{\mu}$ and $\pi_R$ read
	\begin{subequations}\begin{align}\label{96a}
	\nonumber \dot{\pi}_{\mu}=&N\left(2+2\frac{\left(R'\right)^2}{\mu^2}+\frac{1}{4\left(3\lambda-1\right)}
	\left(\frac{\lambda-1}{2}\frac{\pi_R^2}{\mu^2}-\left(2\lambda-1\right)\frac{\pi_{\mu}^2}{R^2}\right)
	\right)\\
	&-4\frac{R'}{\mu^2}\left(N'R+R'N\right)+\xi\pi_{\mu}',\\ \nonumber 
	\dot{\pi}_R=&N\left(\frac{1}{2\left(3\lambda-1\right)}\left(
	\left(2\lambda-1\right)\frac{\mu\pi_{\mu}^2}{R^3}-\lambda\frac{\pi_{\mu}\pi_R}{R^2}\right)-4\frac{R''}{\mu}+4\frac{R'\mu'}{\mu^2}\right)
	\\&-4\frac{R'}{\mu}N'-4\frac{R}{\mu}N''+4\frac{R}{\mu^2}\mu'N'+\xi\pi_R'+\pi_R\xi'.\label{pirdot}
	\end{align}\end{subequations}
Substituting the results for $\pi_{\mu}$, $\pi_R$, $\mu$, $N$ and $\xi$ in terms of $R$ into eq.\ \eqref{96a} reduces it to
	\begin{equation}
	\frac{\dot{C}}{R}+\frac{\dot{A}}{3}R^2=-4\frac{n_2}{R}+\frac{\dot{A}}{3}R^2,
	\end{equation}
which is again solved by $\dot{C}\! =\! -4n_2$. A lengthy algebraic computation shows that eq.\ (\ref{pirdot}) is satisfied if
	\begin{equation}
	\frac{\left(R'\right)^2}{B^2}\left(P_0+P_{-2}R^{-2}+P_{-3}R^{-3}\right)=0,
	\end{equation}
where the $P_k$ are polynomials of degree $k$ in the metric function $R$ and otherwise
functions of $A$, $\dot{A}$ ,$C$, $\dot{C}$, $m$, $\dot{m}$, $n_2$ and $\lambda$. 
Since $\frac{\left(R'\right)^2}{B^2}$ cannot vanish everywhere\footnote{As we will see later, for $A\neq 0$ this combination vanishes in 
the $r\rightarrow\pm\infty$ limit, as a result of which the hypersurface $\Sigma$ becomes asymptotically null.}, 
the individual $P_k(R)$ must vanish identically for all $(r,t)$. 
The condition $P_{-2}\! =\! 0$ is again solved by $\dot{C}\! =\! -4n_2$. 
Once this condition is substituted into both $P_0$ and $P_{-3}$, both equations yield the condition on $\dot{m}$ we already obtained as a 
solution to the equation for $\dot{\mu}$,
	\begin{subequations}\begin{align}
	&P_0=0\Rightarrow A=0 \quad \lor \quad 8\dot{m}=\frac{n_2 A}{3\left(3\lambda-1\right)}
	-\frac{C\dot{A}}{24},\\
	&P_{-3}=0\Rightarrow C=0\quad \lor\quad 8\dot{m}=\frac{n_2 A}{3\left(3\lambda-1\right)}
	-\frac{C\dot{A}}{24}.
	\end{align}\end{subequations}
We conclude that the equations of motion for all phase space variables are solved by the two conditions
	\begin{equation}\label{43}
	8\dot{m}=\frac{n_2 A}{3\left(3\lambda-1\right)}-\frac{C\dot{A}}{24}\quad \land \quad
	\dot{C}=-4n_2.
	\end{equation}
Finally, we should solve $\dot{N}\! =\!\alpha$, with $\alpha$ given by eq.\ \eqref{33}. Expanding $\dot{N}$, we obtain
	\begin{equation}
	\dot{N}=\frac{\partial N}{\partial R}\dot{R}+\frac{\partial N}{\partial n_1}\dot{n}_1+\frac{\partial N}{\partial n_2}\dot{n}_2+\frac{\partial N}{\partial m}\dot{m}+\frac{\partial N}{\partial A}\dot{A}+\frac{\partial N}{\partial C}\dot{C}.
	\end{equation}
Substituting the shift $\xi$ given in eq.\ \eqref{44} into eq.\ \eqref{33} for $\alpha$, the $\dot{R}$-terms 
cancel immediately, as does the $N'$-term in $\alpha$. The same is true for the terms involving $\dot{n}_1$, $\dot{n}_2$ and $\ddot{A}$. 
The remaining terms read
	\begin{equation}
	\alpha=\dot{N}\;\Leftrightarrow\;  b\left(\frac{C}{4R^2}+\frac{AR}{6\left(3\lambda-1\right)}\right)=-\frac{1}{2}\left(\frac{\partial B}{\partial m}\dot{m}+\frac{\partial B}{\partial A}\dot{A}+\frac{\partial B}{\partial C}\dot{C}\right),
	\end{equation}
which is immediately satisfied once eq.\ \eqref{43} is substituted on the right-hand side.

After solving all constraints and equations of motion, only two quantities remain undetermined, the canonical coordinate
$R(t,r)$ and the Lagrange multiplier $\beta(t,r)$ associated with the radial momentum constraint. 
The arbitrary nature of $\beta$ is related to the diffeomorphism symmetry in the radial direction, which the spherically symmetric ansatz
leaves unfixed. 
The associated coordinate freedom can be used to fix $R$ as a function of $(t,r)$. To implement this, one first imposes a gauge-fixing condition
$\xi-\xi_{\it gf}\approx 0$ on the 
shift $\xi$. Demanding that this gauge choice be preserved in time then leads to an equation for $\beta$, namely, 
	\begin{equation}\label{beta}
	\frac{d}{dt}\left(\xi - \xi_{\it gf}\right)=\beta - \dot{\xi}_{\it gf}= \beta - \frac{\partial \xi_{\it gf}}{\partial t} - \poiss{\xi_{\it gf},H_{tot}}\approx 0.
	\end{equation}
Of course, we must make sure that any expression derived for $\xi_{\it gf}$ is compatible with eq.\ \eqref{44}, under the substitution 
$\xi\rightarrow\xi_{\it gf}$.  	
Our gauge choice for $\xi$ is inspired by eq.\ \eqref{44} and reads
	\begin{equation}\label{anz}
	\xi_{\it gf}:=aN_{sol}\left(\frac{C}{4R^2}+\frac{AR}{6\left(3\lambda-1\right)}\right),
	\end{equation}
where $a$ is a real number that will be chosen separately for $r>0$ and $r<0$.
It is an unphysical gauge parameter that is introduced for mere convenience, as will become clear below.
On the constraint surface, $\xi_{\it gf}$ can be written equivalently as
\begin{equation}
\label{xigf}
\xi_{\it gf}=\frac{aN_{sol}}{2\left(3\lambda-1\right)}\left(\lambda\;\!\frac{\pi_{\mu}}{R}+\frac{1-\lambda}{2}\;\frac{\pi_R}{\mu}\right). 
\end{equation}
Substituting expression (\ref{xigf}) into eq.\ \eqref{beta}, the latter becomes
	\begin{equation}\label{beta 2}
	\beta \approx\;\! \frac{\alpha\;\!\xi_{\it gf}}{N_{sol}} + \frac{aN_{sol}}{2\left(3\lambda-1\right)}
	 \poiss{ \left( \lambda\;\!\frac{\pi_{\mu}}{R}+\frac{1-\lambda}{2}\;\frac{\pi_R}{\mu}\right)    ,H_{tot}}.
	\end{equation}
To obtain (\ref{beta 2}), we have used that  
the only contribution to $\frac{\partial \xi_{\it gf}}{\partial t}$ comes from $\frac{\partial N_{sol}}{\partial t}$, which combined with 
$\poiss{N_{sol},H_{tot}}$ yields the $\alpha$-dependent term on the right-hand side of eq.\ \eqref{beta 2}. 
Computing the Poisson bracket in eq.\ \eqref{beta 2}, this equation becomes
	\begin{subequations}\begin{align}
	\beta \approx\; &\frac{\alpha\;\!\xi_{\it gf}}{N_{sol}}+\frac{aN^2_{sol}}{2\left(3\lambda-1\right)}\left(\frac{A^2R}{18\left(3\lambda-1\right)}-\frac{3\lambda-1}{4}\frac{C^2}{R^5}-\frac{AC}{12R^2}\right)\left(aR'-1\right)
	\\&+aN_{sol}\left(\frac{\dot{A}R}{6\left(3\lambda-1\right)}-\frac{n_2}{R^2}\right).
	\end{align}\end{subequations}
Computing $\dot{\xi}_{\it gf}$ and substituting it into eq.\ (\ref{beta}), $\beta - \dot{\xi}_{\it gf}\!\approx\! 0$, together with the expression 
just obtained for $\beta$ yields
	\begin{subequations}\begin{align}
	\beta-\dot{\xi}_{\it gf}\approx\; &\frac{aN^2_{sol}}{2\left(3\lambda-1\right)}\left(\frac{A^2R}{18\left(3\lambda-1\right)}-\frac{3\lambda-1}{4}\frac{C^2}{R^5}-\frac{AC}{12R^2}\right)\left(aR'-1\right)
	\\&-aN_{sol}\dot{R}\left(\frac{A}{6\left(3\lambda-1\right)}-\frac{C}{2R^3}\right)\approx 0,
	\end{align}\end{subequations}
which is solved by $\dot{R}\! =\! 0$ and $R'\! =\!\frac{1}{a}$. These are precisely the same conditions as one obtains from demanding consistency
between eqs.\ \eqref{anz} and \eqref{44}. 

In the remainder of the paper, we will set $\dot{R}=0$ but not fix $R$ as function of the coordinate $r$, to emphasize the validity of our results 
for general $R(r)$. The only exception will be the discussion of boundary conditions in Sec.\ \ref{fall}, where we will set $R=\left|r\right|$.
Furthermore, we will choose $a\! =\! 1$ for $r\! >\! 0$ and $a\! =\!-1$ for $r\!<\! 0$. 
The motivation for this choice is to have the same spacetime for positive and negative $r$. 
As can be seen from the definition of $B$ in terms of $R$, $R$ must be even with respect 
to the inversion $r\rightarrow -r$ for this to happen. Moreover, this choice is needed to have the integrand in the lapse vanish 
for $r\rightarrow-\infty$, implying that $n_1$ determines the behaviour of the lapse at both spatial infinities.

Note that setting $\dot{R}=0$ does not remove all time dependence from the metric. This would only be true if
$\dot{A}$, $\dot{C}$, $\dot{n}_1$ and $\dot{m}$ vanished too, which would imply a considerable restriction on the space of solutions. 
However, we can still use conditions \eqref{43} to define a quantity $M$ 
that is conserved, $\dot{M}=0$, and in such a way that $B$ contains a term of the form $1-\frac{2M}{R}$. 
For the general relativistic case $\lambda=1$ this is achieved in a straightforward manner by noting that \eqref{43} simplifies to
	\begin{equation}
	8\dot{m}=-\frac{1}{24}\left(\dot{C}A+C\dot{A}\right)
	\end{equation}
which implies that
\begin{equation}
2M=8m+\frac{C A}{24}
\end{equation}	
is conserved. For the general case $\lambda\neq 1$, we define the conserved $M$ by
	\begin{equation}\label{M}
	2M:=8m+\frac{C A}{12\left(3\lambda-1\right)}+\frac{\lambda-1}{8\left(3\lambda-1\right)}\int_{-\infty}^tdt' C\dot{A},
	\end{equation}
where we have set the lower integration limit to $-\infty$ to have $\dot{M}(t)\! =\! 0$ for all times $t$.
In order for the integral to exist and be finite, we must demand that the functions $A(t)$ and $C(t)$ are such that 
$C\dot{A}$ goes to zero faster than $1/t$ in the limit $t\rightarrow -\infty$. We will assume in the following that this is the case. 
It allows us to rewrite the function $B$ as
	\begin{equation}
	B=1-\frac{2M}{R}+\frac{1}{3\lambda-1}\left(\frac{CA}{12}+\frac{\lambda-1}{8}\int_{-\infty}^t \!\! d\tilde{t}\;
	C\dot{A}\right)\frac{1}{R} +\frac{C^2}{16 R^4}+\frac{A^2R^2}{72\left(3\lambda-1\right)}.\label{Blong}
	\end{equation}
Before turning to the discussion of the model's solutions and their properties, let us finally substitute $n_2=-{\dot C}/4$ into the 
lapse function (\ref{36}) and set $r_0=+\infty$, yielding
	\begin{equation}
	N_{sol}=\sqrt{B}\left(n_1+\frac{1}{4}\int_{r}^{\infty}d\tilde{r}\frac{R'}{B^{3/2}}\left(\frac{\dot{C}}{R^2}+\frac{\dot{A}R}{3}\right)\right).
	\label{Nnew}
	\end{equation}
Inspecting (\ref{Nnew}), we can reconfirm that the function $n_1(t)$ determines the value of the lapse at radial infinity, 
as stated earlier below eq.\ (\ref{lap}).	

\section{The $\lambda$-dependent spherically symmetric solutions}
\label{prop}
Our next step will be to determine some geometric properties of the solutions we have obtained. Computing the extrinsic
curvature of the constant-time slices will lead to a geometric interpretation of the functions $A(t)$ and $C(t)$ introduced
earlier. This will be followed by a discussion of boundary and fall-off conditions that must be imposed on the fields. 
Implementing them enables us to write the four-dimensional metrics corresponding to the $\lambda$-R solutions in a form where
they can be compared easily with their general relativistic counterparts.
Finally, we compute the four-dimensional scalar curvature $^{(4)}\! R$ of the $\lambda$-R model and find it to be 
proportional to $(\lambda\! -\! 1)$ and
nonvanishing as long as the trace of the extrinsic curvature does not vanish, a situation which is very different from that 
in standard gravity.

\subsection{Extrinsic curvatures}
\label{princ}
We begin by re-expressing the extrinsic curvatures of (\ref{Krr}), (\ref{Kpp}) in terms of the parameters of the
reduced phase space,
	\begin{subequations}\label{K}\begin{align}
	K_{rr}=\; &\mu^2\left(\frac{C}{2R^3}-\frac{A}{6\left(3\lambda-1\right)}\right),\\
	K_{\theta\theta}=\; & \frac{K_{\phi\phi}}{\sin^2\theta}= -\frac{AR^2}{6\left(3\lambda-1\right)}-\frac{C}{4R}.
	\end{align}\end{subequations}
From this, the trace of the extrinsic curvature $K=g^{ij}K_{ij}$ (the ``mean curvature") can be computed 
straightforwardly and, up to a $\lambda$-dependent constant, turns out to be equal to the integration constant $A(t)$ first 
introduced in eq.\ (\ref{17}) above,
\begin{equation}
K=-\frac{A}{2\left(3\lambda-1\right)}\;\; \Rightarrow\;\; A=-2\left(3\lambda-1\right)K.\label{k1}
\end{equation}
We have therefore shown that the mean curvature of the slices of the preferred foliation is spatially constant.
Using \eqref{K}, we can now also justify our previous assertion that $C$ measures the transverse-traceless components 
of the extrinsic curvature $K_{ij}$. Defining the traceless extrinsic curvature tensor $K_{ij}^T$ by
	\begin{equation}\label{TT}
	K_{ij}^T:=K_{ij}-\frac{1}{3}g_{ij}K,
	\end{equation}
the principal curvatures $K^T\!{}_i{}^i$ -- the coordinate-independent eigenvalues of the Weingarten map -- 
are found to be
\begin{equation}
{K^T\!}_r{}^r=\frac{C}{2 R^3}\, ,\;\;\;\;\; K^T\!{}_\theta{}^\theta=K^T\!{}_\phi{}^\phi=-\frac{C}{4 R^3}.
\label{princip}
\end{equation}
The fact that they only depend on $C$ and $R$
shows that $C$ carries all the transverse-traceless information of the extrinsic curvature. 

To be able to compare our results with the general CMC foliations of the Schwarzschild geometry, 
we now introduce the same variables as in \cite{CMCSch} and replace $A$ by $K$ everywhere, leading to
	\begin{eqnarray}\label{66}
	\frac{\mu^2}{\left(R'\right)^2}\!\!\! & =&\!\!\! \frac{1}{B}\!=\!\bigg(1-\frac{2M}{R}+\!\left(\frac{K R}{3} - \frac{C}{4R^2}\right)^2\!\!
	+\!\!\left(\lambda-1\right)\!\left(\frac{K^2R^2}{6}-\frac{1}{4R}\int_{-\infty}^t\!\! dt' C\dot{K}\right)\!\!\! \bigg)^{-1},\nonumber\\
	N \!\!&=&\!\! \sqrt{B}\,\bigg(n_1+\frac{1}{4}\int_r^{\infty}\!\! d\tilde{r}\;\frac{R(\tilde{r})'}{B^{3/2}}\bigg(\frac{\dot{C}}{R(\tilde{r})^2}-\frac{4}{3}\dot{K}R(\tilde{r})-2\left(\lambda-1\right)\dot{K}R(\tilde{r})\bigg)\!\!\bigg),\nonumber\\
	\xi\!\! &=&\!\! \frac{N}{R'}\bigg(\frac{C}{4R^2}-\frac{KR}{3}\bigg).
	\end{eqnarray}

\subsection{Fall-off conditions and boundary Hamiltonian}
\label{fall}

As mentioned in Sec.\ \ref{red mod}, the Hamiltonian $H$ in general has to include a boundary 
term $H_{\delta\Sigma}$ to make the variational principle well defined, in the sense that its variation $\delta H$ 
can be written as 
	\begin{equation}\label{variation}
	\delta H=\int d^3x\left(A^{ij}\delta g_{ij}+B_{ij}\delta \pi^{ij}\right),
	\end{equation}
without any boundary contributions on the right-hand side, such that the equations of motion
	\begin{equation}\label{e.o.m.}
	\dot{g}_{ij}=\frac{\delta H}{\delta \pi^{ij}}\equiv B_{ij},\qquad \dot{\pi}^{ij}=-\frac{\delta H}{\delta g_{ij}}\equiv A^{ij}
	\end{equation}
follow from it in a unique manner \cite{RgTt} (see also the related discussion in \cite{KK}).
In our reduced setting, equation \eqref{variation} becomes
	\begin{equation}\label{red variation}
	\delta H=\int_{-\infty}^{+\infty}dr\left(A_{\mu}\delta\mu+A_R\delta R+B_{\mu}\delta\pi_{\mu}+B_R\delta\pi_R\right).
	\end{equation}
A straightforward variation of the Hamiltonian \eqref{10} does not yield an equation of the form \eqref{red variation},
because some phase space variables occur in the Hamiltonian with spatial derivatives.
To cast their variation into a form matching the terms in the integrand of \eqref{red variation}, one has to perform partial integrations, leading
to additional boundary terms. We will collect all boundary contributions generated in this way, impose the coordinate
condition $R=|r|$ motivated in Sec.\ \ref{tev} above, evaluate the boundary terms on solutions,
and finally determine the boundary Hamiltonian whose variation cancels these unwanted contributions.

Note that we have a nonvanishing boundary Hamiltonian from the outset, because a partial integration 
	\begin{equation}\label{14}
	\int_{-\infty}^{+\infty}dr \pi_{\mu}\left(\mu\xi\right)'=-\int_{-\infty}^{+\infty}dr
	\mu\xi\pi_{\mu}'+\left.\mu\xi\pi_{\mu}\right|_{-\infty}^{+\infty}
	\end{equation}
has to be performed to obtain the Hamiltonian $H$ in the form \eqref{10}. In addition, we find a boundary contribution 
	\begin{equation}\label{lim1}
	\left. \xi (\pi_R\, \delta R -\mu\,\delta\pi_{\mu})\right|_{-\infty}^{+\infty}
	\end{equation}
from the variation of the shift-dependent term in \eqref{10} and a contribution
	\begin{equation}\label{lim2}
	4\, \bigg(\left. \!\frac{NR}{\mu}\, \delta R' -\frac{N'R}{\mu}\,\delta R-\frac{NRR'}{\mu^2}\,\delta\mu\bigg)\right|_{-\infty}^{+\infty}
	\end{equation}
from varying the lapse-dependent term.
Adding \eqref{14} and \eqref{lim1}, the shift-dependent boundary variation is given by
	\begin{equation}\label{lim3}
	\big( \left. \xi\pi_R\, \delta R +\pi_{\mu}\, \delta (\xi\mu)\big) \right|_{-\infty}^{+\infty}.
	\end{equation}
We now implement the gauge-fixing $R\! =\!\left|r\right|$, which implies $R'\! =\! -1$ for $r<0$ and $R'\! =\! 1$ for $r > 0$. In line with our earlier
comments below eq.\ \eqref{Nsol} this means that for $r>0$ we have $\mu\! =\! R'/\sqrt{B}$, while for $r<0$ we must use
$\mu\! =\! -R'/\sqrt{B}$, leading to $\mu\! =\! B^{-1/2}$ for either sign of the coordinate $r$.
With this choice, both $\delta R$ and $\delta R'$ vanish. Setting the corresponding terms in the variations to zero, 
we now substitute the solutions for $\mu$, $R$, $N$, $\xi$ and $\pi_{\mu}$ in terms of integration constants into the remainders.
Expression \eqref{lim2} becomes
	\begin{eqnarray}\label{lim1ex}
	4\left.NRR'\, \delta\frac{1}{\mu}\right|_{-\infty}^{+\infty}\!\!  &=&\!\!  2\left.n_1\left|r\right|\left|r\right|'\delta\!\left(1-\frac{8m}{\left|r\right|}+\frac{C^2}{16r^4}+\frac{3\lambda-1}{18}K^2r^2\right)\right|_{-\infty}^{+\infty}\nonumber \\
	&=&\!\!  2\lim_{r\rightarrow\infty}n_1\left(\frac{2}{9}\left(3\lambda-1\right)\left|r\right|^3 K\delta K-16\,\delta m\right),
	\end{eqnarray}
while expression \eqref{lim3} yields
	\begin{eqnarray}\label{lim3ex}
	\left.\pi_{\mu}\, \delta (\xi\mu)\right|_{-\infty}^{+\infty}\!\!  &=& \!\!\! 
	\left.\left(\frac{C}{\left|r\right|}-\frac{2}{3}\left(3\lambda-1\right)Kr^2\right)\frac{1}{\left|r\right|'}\,
	\delta\! \left[n_1\left(\frac{C}{4r^2}-\frac{K\left|r\right|}{3}\right)\right]\right|_{-\infty}^{+\infty} \nonumber
	\\&=&\!\! 
	2\lim_{r\rightarrow\infty}\left\{\delta n_1\left(\frac{2}{9}\left(3\lambda-1\right)K^2\left|r\right|^3-\frac{3\lambda+1}{6}CK\right)\right. \nonumber\\
	&&\!\! +\left.n_1\left(\frac{2}{9}\left(3\lambda-1\right)\left|r\right|^3K\delta K-\frac{\delta\left(KC\right)}{3}+\frac{1-\lambda}{2}K\delta C\right)
	\right\}.
	\end{eqnarray}
We first consider the last term of expression \eqref{lim1ex},
	\begin{equation}
	\label{var m} 
	- 32\lim_{r\rightarrow\infty} n_1\;\! \delta m =- 32\, n_1\;\! \delta m,
	\end{equation}
which only depends on time, because both $m$ and $n_1$ are spatially constant. To make this vanish, we could demand that
$n_1(t)=0$. This would imply that the lapse vanishes at spatial infinity and that no time evolution takes place there, which is not physically 
acceptable.

Alternatively, we can include a term
$32\, n_1m$ in the boundary Hamiltonian $H_{\partial\Sigma}$, as a consequence of which we would have to demand that 
\begin{equation}
32\, m\,\delta n_1=0.
\label{altern}
\end{equation}
However, setting $m\! =\! 0$ would imply $M\! =\! 0$ in the asymptotically flat case, 
as follows straightforwardly from eq.\ \eqref{M} when the convergence condition for the
integrand $C\dot A$ are satisfied, as we are assuming. This condition appears too restrictive, since it would not even allow for the standard
Schwarzschild solution. 

The only other possibility to satisfy eq.\ \eqref{altern}, arguing along the lines of \cite{KK}, is to assume that 
$n_1$ is a prescribed function at radial infinity (and thus everywhere), which we therefore do not vary. 
Adopting this assumption and setting $\delta n_1=0$, we can add the remaining nonvanishing variations from expressions
\eqref{lim1ex} and \eqref{lim3ex}, leading to 
	\begin{equation}\label{imp var}
	2\lim_{r\rightarrow\infty}n_1\left(  \frac{2}{9}\left(3\lambda-1\right)\left|r\right|^3\delta\! \left(K^2\right)
	- \frac{\delta\left(KC\right)}{3}+\frac{1-\lambda}{2}\, K\delta C 
	\right).
	\end{equation}
If we allow arbitrary variations of $m$, $K$ and $C$, it is impossible to write down a boundary Hamiltonian whose variation 
would cancel all terms in \eqref{imp var}. One obstruction is the term proportional to $(1-\lambda)$, because it cannot be
written as a total variation. A second issue is that the boundary term necessary to cancel the term proportional to
$\delta (K^2)$ in \eqref{imp var} is manifestly divergent. Both of these issues are resolved by setting $\delta K$ to zero at radial infinities,
$\left.\delta K\right|_{\left|r\right|\rightarrow\infty}\! =\! 0$, and therefore everywhere. 
Together with the condition $\left.\delta n_1\right|_{\left|r\right|\rightarrow\infty}\! =\! 0$ this implies 
$\left.\delta N\right|_{\left|r\right|\rightarrow\infty}\! =\! 0$, as can be seen by inspecting eqs.\ \eqref{Blong} and \eqref{Nnew}.
Taking all of these considerations into account, we arrive at a finite expression for the boundary Hamiltonian, namely
	\begin{equation}\label{bound H}
	H_{\delta\Sigma}=n_1\left( 32\;\! m -\frac{3\lambda-1}{3}K C\right),
	\end{equation}
accompanied by the conditions
\begin{equation}
\delta K=0,\;\;\; \delta n_1=0.
\label{cond}
\end{equation}
Note that in the general relativistic case $\lambda\! =\! 1$, $K\! =\! C\! =\! 0$, expression (\ref{bound H}) coincides with the one given in \cite{KK}.

There is one subtlety we have so far not spelled out explicitly in our discussion of the boundary Hamiltonian.
Since our spatial coordinate system is not well-defined for $r=0$, we have been working implicitly with two distinct coordinate patches 
for every spatial hypersurface, defined by $r>0$ and $r<0$. 
However, there is no reason why the integration constants chosen for both patches should be the same.
For full generality the set of constants should be twice as large, for example, $m$ should be replaced by $m_+$ for $r>0$ and $m_-$ for $r<0$. 
Doubling all constants in this manner leads to a boundary Hamiltonian of the form
	\begin{equation}\label{bound H fin}
	H_{\delta\Sigma}=\lim_{r\rightarrow+\infty} n_{1+}\left[16m_+-\frac{3\lambda-1}{6}K_+C_+\right]+\lim_{r\rightarrow-\infty} n_{1-}\left[16m_--\frac{3\lambda-1}{6}K_-C_-\right],
	\end{equation}	
with conditions \eqref{cond} replaced by 
	\begin{equation}
\delta K_\pm =0,\;\;\; \delta n_{1\pm}=0.	
	\end{equation}
Although we will in the remainder of the text not distinguish between integration constants for the positive and the negative $r$-patch, 
it should be understood that there is in principle one set of distinct constants for each.

As a final comment, note that while $g_{rr}$ vanishes as $r\rightarrow\pm\infty$, the vector $\partial_t$ does not become null in this limit,
as can be seen by computing $g_{00}\equiv N^iN_i-N^2$, leading to
	\begin{equation}
	\lim_{r\rightarrow\pm\infty}\left(N^iN_i-N^2\right)=\lim_{r\rightarrow\pm\infty}\left(\mu^2\xi^2-N^2\right)=\frac{n_1^2A^2r^2}{24\left(3\lambda-1\right)^2}\left(1-\lambda\right),
	\label{gzero}
	\end{equation}
and implying that $\partial_t$ is timelike for $\lambda>1$ and spacelike for $\lambda<1$.\footnote{The case $\lambda =1$ must be considered 
separately; the leading term on the right-hand side of eq.\ (\ref{gzero}) in this case is of order $r^0$ and negative, implying a timelike vector
$\partial_t$.} The fact that the vector $\partial_t$ associated with the time coordinate $t$ 
can become spacelike when the shift is large is a function of the foliation and a familiar feature,
for example, from the Painlev\'e-Gullstrand representation of the Schwarzschild metric inside the event horizon.  
It illustrates how different values of the parameter $\lambda$ can affect aspects of the foliation structure. 
While the time vector $\partial_t$ can cease to be timelike, 
the normal evolution vector $\vec{m}=\vec{n}N$, with $\vec{n}$ the unit normal to the hypersurface, will of course remain timelike whenever 
the hypersurface is spacelike (or null, when the hypersurface is null).

\subsection{Four-dimensional metric}
\label{4dimg}
Using the expressions \eqref{66} for $\mu^2$, $N$ and $\xi$, we can write the $g_{0\mu}$-components of the four-dimensional 
metric of the solutions of the $\lambda$-R model as
	\begin{subequations}\begin{align}
	\nonumber
	g_{00}=&-N^2+\mu^2\xi^2=-\frac{N^2}{B}\left(B-\left(\frac{C}{4R^2}-\frac{KR}{3}\right)^2\right)
	\\=&-\frac{N^2}{B}\left(1-\frac{2M}{R}+\left(\lambda-1\right)\left(\frac{K^2R^2}{6}-\frac{1}{4R}\int_{-\infty}^t \!\!\! dt'\, C\dot{K}\right)\right),
	\label{g00}\\
	\nonumber
	g_{0r}=&\;\mu^2\xi=\frac{R'N}{B}\left(\frac{C}{4R^2}-\frac{KR}{3}\right)
	=\frac{R'}{\sqrt{B}}\left(\frac{C}{4R^2}-\frac{KR}{3}\right)\!\!\\
	&\times \bigg(n_1+\frac{1}{4}\int_r^{\infty}\!\!\!
	d\tilde{r}\;\frac{R(\tilde{r})'}{B^{3/2}}\bigg(\frac{\dot{C}}{R(\tilde{r})^2}-\frac{4}{3}\,\dot{K}R(\tilde{r})-
	2\left(\lambda-1\right)\dot{K}R(\tilde{r})\bigg)\!\!\bigg),\label{g0r}
	\end{align}\end{subequations}
where the quotient $\frac{N^2}{B}$ in (\ref{g00}) is given by
	\begin{equation}
	\frac{N^2}{B}=n_1+\frac{1}{4}\int_r^{\infty}\!\! d\tilde{r}\;\frac{R(\tilde{r})'}{B^{3/2}}\bigg(\frac{\dot{C}}{R(\tilde{r})^2}-\frac{4}{3}\,\dot{K}R(\tilde{r})-2\left(\lambda-1\right)\dot{K}R(\tilde{r})\bigg).\label{n2b}
	\end{equation}
The $g_{rr}$-entry of the metric is given by $\mu^2=(R')^2/B$, which was given in the first relation of (\ref{66}). It is straightforward 
to show that $g_{rr}$ goes to zero as $|r|\rightarrow\infty$, which implies that the hypersurfaces of constant time become asymptotically
null, as we have already stated on several occasions.
For the inverse metric, we find
	\begin{subequations}\begin{align}
	&g^{00}=-\frac{1}{N^2},\qquad g^{0r}=\frac{\xi}{N^2}=\frac{1}{R'N}\left(\frac{C}{4R^2}-\frac{KR}{3}\right),\\
	&g^{rr}=\frac{1}{\mu^2}-\frac{\xi^2}{N^2}=\frac{1}{\left(R'\right)^2}\left(1-\frac{2M}{R}+\left(\lambda-1\right)
	\left(\frac{K^2R^2}{6}-\frac{1}{4R}\int_{-\infty}^t dt' C\dot{K}\right)\!\!\right).\label{71b}
	\end{align}
	\end{subequations}
Summarizing, the four-dimensional $g_{\mu\nu}$ and its inverse $g^{\mu\nu}$ are given by
	\begin{subequations}\label{72}\begin{align}
	g_{\mu\nu}=&\left(\begin{array}{cccc}
	-\frac{N^2}{B}\left(B-\left(\frac{C}{4R^2}-\frac{KR}{3}\right)^2\right) & \frac{R'N}{B}\left(\frac{C}{4R^2}-\frac{K R}{3}\right) & 0 & 0 \\ 
	\frac{R'N}{B}\left(\frac{C}{4R^2}-\frac{K R}{3}\right)  & \frac{\left(R'\right)^2}{B} & 0 & 0 \\
	0 & 0 & R^2 & 0 \\ 0 & 0 & 0 & R^2\sin^2\theta
	\end{array}\right),\label{gmunu}\\
	g^{\mu\nu}=&\left(\begin{array}{cccc}\vspace{0.2cm}
	-\frac{1}{N^2} &\frac{1}{R'N}\left(\frac{C}{4R^2}-\frac{KR}{3}\right) & 0 & 0\\
	\frac{1}{N}\left(\frac{C}{4R^2}-\frac{KR}{3}\right) & 
	\frac{1}{\left(R'\right)^2}\left(B-\left(\frac{C}{4R^2}-\frac{KR}{3}\right)^2\right) & 0 & 0\\
	0 & 0 & \frac{1}{R^2} & 0 \\ 0 & 0 &0 & \frac{1}{R^2\sin^2\theta}
	\end{array}\right).
	\end{align}\end{subequations}
We would like to compare the expressions for the entries of the four-metric to their counterparts in the CMC-description of general
relativity in reference \cite{CMCSch}.\footnote{There is a discrepancy between our result \eqref{71b} for $g^{rr}$ (for $\lambda\! =\! 1$)
and that of \cite{CMCSch}, apparently because $g^{ij}$ was used instead of the correct four-dimensional inverse 
$^{(4)}g^{ij}\! =\! g^{ij}\! -\! \frac{N^iN^j}{N^2}$.} 
Rewriting our results in terms of $K$ and isolating the $\lambda$-dependence into terms proportional to $(\lambda - 1)$ in 
expressions (\ref{g00}), (\ref{g0r}), (\ref{n2b}) and (\ref{71b}) has made explicit how the spacetime metric 
$g_{\mu\nu}$ differs from its general relativistic counterpart. 
As we will show in Sec.\ \ref{4dimR}, these extra contributions lead to a nonvanishing four-dimensional curvature for $K\neq 0$ and $\lambda\neq 1$.

The four-dimensional metric we have derived depends on five parameters, two constants ($\lambda$, $M$) and 
three functions of time ($C$, $K$, $n_1$). Let us comment on their role and interpretation in turn. 
The coupling constant $\lambda$ only occurs in the prefactors $\left(\lambda-1\right)$ of terms that do not appear in the Schwarzschild solution. 
The constant $M$ was defined in eq.\ \eqref{M} from the integration constant $m(t)$, which was introduced earlier when solving the Hamiltonian
constraint, to have a genuinely conserved quantity that reduces to a multiple of the Schwarzschild mass $M_s$ for $\lambda\! =\! 1$. 
It can be checked that in this latter case we have $M\! =\! M_s/4$ with our choice of units. 

When $\lambda\! =\! 1$, neither $C$ nor $K$ play a direct physical role. However, they determine the range of $R$ for which the function $B(R)$ 
is positive, which in turn determines the spacetime covered by the slices of the foliation. 
More specifically, as was shown in \cite{CMCSch1,CMCSch2} for $\dot{K}\! =\! 0$ and later in \cite{CMCSch} for $\dot{K}\!\neq\! 0$, 
the number and location of the roots of $B$ depends on both parameters. 
Keeping $K\! >\! 0$ fixed, there are three possibilities. If $C\! =\! 0$, $B$ has only one root. In this case, the foliation extends from null infinity to this 
minimal radius, re-emerges on the other side of the ``throat" and continues from there all the way to the other null infinity. 
For small $C\! >\! 0$, there are two roots and two regions for which $B$ is positive, one in the interior black hole region of the Kruskal
diagram, extending from the singularity $R=0$ to some maximal radius and then returning to the singularity, 
and another one retaining the $C\! =\! 0$ behaviour. 
Lastly, if $C$ is large enough, there is a critical point for which these two roots coincide, beyond which the leaves of the foliation start 
at either of the null infinities and end again in the singularity. 
We expect a qualitatively similar behaviour in our solutions, certainly for small deviations from the general relativistic case,
although the roots of $B$ will of course become $\lambda$-dependent. 

Regarding the role played by $C$ and $K$ in the $\lambda$-R model, recall that the former is obtained when solving the first-class radial 
momentum constraint, while the latter is associated with the second-class tertiary constraint $\omega\approx 0$. 
This could suggest that $K$ is a physical quantity while $C$ is not, but the argument turns out to be more involved.
Unlike what happens in general relativity, the lapse function $N$ is not determined by making a gauge choice  
but by solving the quaternary constraint ${\cal M}\!\approx\! 0$, and $N$ depends on both $C$ and $n_1$. 
We will show in the next section that $K$ is a physical quantity, in the sense that the four-dimensional scalar curvature -- a scalar under
local diffeomorphisms --  depends on it. 
However, a similar logic applies to $C$ and $n_1$, by virtue of their appearance in the lapse function: changing either $C$ or $n_1$ while
keeping all other parameters fixed will alter the lapse and consequently yield a different four-dimensional Ricci scalar. 
At the same time, $C$ and $n_1$ retain the interpretation they had in general relativity, namely, as 
the transverse-traceless part of the extrinsic curvature and the leading-order behaviour of the lapse at infinity 
when $K\neq 0$ and $\lambda\neq 1$.

\subsection{Four-dimensional scalar curvature}
\label{4dimR}
A direct way to obtain the four-dimensional Ricci scalar $\!\!^{(4)}\! R$ is to perform a full, four-dimensional calculation starting from
the explicit expression (\ref{gmunu}) of the four-metric.
We will instead use an expression for $\!\!^{(4)}\! R$ in terms of three-dimensional quantities, which can be derived by
combining the contracted Ricci and Gauss equations \cite{num}.\footnote{Note that the sign of the term linear 
in $K$ on the right-hand side of (\ref{GC}) is opposite to that given by Gourgoulhon \cite{num}, because his extrinsic curvature has 
the opposite sign of ours.} It reads
	\begin{equation}\label{GC}
	^{(4)}R={\cal R}+K^2+K_{ij}K^{ij}+\frac{2}{N}{\cal L}_{N\vec{n}}\;\! K-\frac{2}{N}g^{ij}\nabla_i\nabla_jN,
	\end{equation}
where ${\cal L}_{N\vec{n}}$ is the Lie derivative along the normal evolution vector $N\vec{n}$, 
and $\vec{n}$ the unit normal to the hypersurface $\Sigma$,
	\begin{equation}
	\vec{n}=N^{-1}\left(1,-N^i\right).
	\end{equation}
We first substitute the solutions obtained for the phase space variables into the expression (\ref{intrR}) for 
the scalar three-curvature, resulting in
	\begin{equation}
	{\cal R}=\
	\frac{2}{R^2}\left(1-B-R\;\frac{\partial B}{\partial R}\right).
	\end{equation}
The term with the Lie derivative is given by
	\begin{equation}
	\frac{2}{N}{\cal L}_{N\vec{n}}\;\! K=\frac{2\dot{K}}{N},
	\end{equation}
while the $K_{ij}K^{ij}$-term can in a straightforward way be obtained from eqs.\ \eqref{K},
	\begin{equation}
	K_{ij}K^{ij}=\frac{3}{8}\frac{C^2}{R^6}+\frac{K^2}{3}.
	\end{equation}
To determine the last term in eq.\ (\ref{GC}), we recall the form of the lapse $N$ given in \eqref{66} as a function of $\dot{K}$ and $R$,
and compute its Laplacian as	
	\begin{equation}
	-\frac{2}{N}\nabla_i\nabla^iN=-\frac{3\lambda-1}{N}\dot{K}-\left(\frac{\partial^2 B}{\partial R^2}+\frac{2}{R}\frac{\partial B}{\partial R}\right).
	\end{equation}
Combining all contributions finally yields the four-dimensional scalar curvature
	\begin{eqnarray}\label{curv}
	^{(4)}R\!\!\! &=&\!\!\! -3\left(\lambda-1\right)\frac{\dot{K}}{N}+\frac{3}{8}\frac{C^2}{R^6}+\frac{4K^2}{3}+
	\frac{2}{R^2}\left(1-B-2R\frac{\partial B}{\partial R}-\frac{R^2}{2}\frac{\partial^2 B}{\partial R^2}\right)\nonumber\\
	&=&\!\! (1-\lambda) \bigg( 2K^2+\frac{3\dot{K}}{N}\bigg).
	\end{eqnarray}
This expression vanishes in the general relativistic case $\lambda\! =\! 1$, as it should, and also for vanishing mean curvature, $K\! =\! 0$.
If neither $\lambda\! =\! 1$ nor $K\! =\! 0$, it is necessarily nonzero, because the nontrivial radial dependence of the shift
($N'\neq 0$) prevents a tuning of the initial data $K$ and $\dot{K}$ such that $\!^{(4)}\! R$ vanishes.

\section{Summary and conclusions}
\label{conclusion}

We have succeeded in finding the general solution to the $\lambda$-R model for the case that its preferred spatial hypersurfaces 
possess spherical symmetry. As anticipated, we do not have an analogue of Birkhoff's theorem, since the solutions are
in general non-flat, non-static, incompatible with asymptotic flatness 
and parametrized not only by their conserved mass $M$, but also by the mean extrinsic curvature $K(t)$ of the leaves of the foliation,
as well as prescribed functions $C(t)$ and $n_1(t)$.

Because of the restriction to foliation-preserving diffeomorphisms, our general solutions have a nonvanishing
radial shift, which cannot be removed by an allowed diffeomorphism. 
In agreement with the full $\lambda$-R model, only constant mean curvature and maximal slicings are permitted by the dynamics. 
Solving the (second-class) constraint algebra, and imposing fall-off conditions and time evolution equations, 
we have derived the explicit functional form $g_{\mu\nu}$ of the general spherically symmetric solution of
$\lambda$-R gravity of the class considered, given in eqs.\ \eqref{72}. 

The $\lambda$-dependent constant mean curvature solutions ($K\not\! =\! 0$) are {\it not} physically equivalent 
to the ones with maximal slicing ($K\! =\! 0$). Moreover, only the latter correspond to vacuum solutions of general relativity, 
as follows from the nonvanishing of the four-dimensional Ricci scalar $\! {}^{(4)}\! R$ of eq.\ \eqref{curv} 
in the CMC case. 
Like in general relativity, the Ricci scalar is of course a local invariant. That it can be nonvanishing for $\lambda\not= 1$, 
even in the absence of matter, has to do with the fact that the $\lambda$-R model possesses a local invariant {\it not}
present in general relativity, namely, the trace $K$ of the extrinsic curvature of the distinguished foliation. 

In general relativity, the CMC foliations of the Schwarzschild geometry are all equivalent and can be obtained from 
the usual asymptotically flat metric description (with $K\! =\! 0$ and $C\! =\! 0$) by means of space-dependent time 
reparametrizations \cite{CMCSch1}. 
While these diffeomorphisms generate nonvanishing values for $K$ and $C$, they do not change the geometry of the spacetime, 
but only the way in which the $3+1$ split is implemented. Thus $K$ and $C$ can be thought of as unphysical gauge parameters.
By contrast, the same transformations are no longer allowed symmetries of the $\lambda$-R model, and spacetimes related
by them will in general correspond to physically inequivalent solutions. For each $\lambda >1/3$, $\lambda\not= 1$, the function $K(t)$ 
becomes effectively physical and parametrizes physically distinct spacetimes, as is clear from the functional form of 
the scalar curvature \eqref{curv} in terms of $K(t)$. 
Although the standard, general relativistic solution is included among those of the $\lambda$-R model 
(namely, for initial data $K\! =\! 0$),
it is even for $\lambda\neq 1$ not unique and as a consequence of the preferred foliation can only be attained in a restricted set of coordinate charts.

As in previous work on the $\lambda$-R model (see \cite{LP} and references therein), an interesting question is to what extent 
physical observables are sensitive to the presence of the parameter $\lambda$. An obvious thing to attempt 
in the presence of spherical symmetry is a quantitative re-evaluation of classic solar system tests of general relativity, 
like light deflection or perehelion precession, in order to understand what observational bounds exist on 
deviations of $\lambda$ from its canonical value of 1, and also to quantify the influence of different choices of $K(t)$,
$C(t)$ and $n_1(t)$. 
Of course it should be kept in mind that we have only discussed the specific case where the $SO(3)$-orbits of the
spherical symmetry are aligned with the preferred foliation. This will not be the case for a general spherically symmetric solution
of the $\lambda$-R model, whose treatment will be technically more involved. A convenient framework to tackle this problem may be
the so-called covariant 1+1+2 formalism (see, e.g.\ \cite{cb}), which in addition to a preferred time direction uses a preferred
spatial direction, which in our case would be given by the radial direction perpendicular to the shells of spherical symmetry.

Another possible generalization concerns the inclusion of a cosmological constant term in the action, which we do not expect
to present any difficulties. Given the way in which the cosmological constant $\Lambda$ appears in 
the $\lambda$-dependent version of the Lichnerowicz-York equation \cite{LN} and in the usual 
Schwarzschild-de Sitter solution, we anticipate that this will lead to a $\Lambda$-dependent version of the term proportional 
to $R^2$ in the function $B(R)$ of eqs.\ (\ref{bform}) and (\ref{66}).

\vspace{.5cm}

\noindent {\bf Acknowledgements.} 
LP acknowledges financial support from Funda\c{c}\~{a}o para a Ci\^{e}ncia e Tecnologia, Portugal through 
grant no. SFRH//BD/76630/2011 and thanks S. Gryb for helpful discussions during various stages of this work.

\end{document}